\documentstyle[12pt,cite]{article}
\newcommand{\infig}[2]{\begin{center}\mbox{ \epsfxsize #1
                       \epsfbox{#2}}\end{center}}

\newcommand{\be}{\begin{equation}}
\newcommand{\nn}{\nonumber}
\newcommand{\ee}{\end{equation}}
\newcommand{\bea}{\begin{eqnarray}}
\newcommand{\eea}{\end{eqnarray}}

\input epsf
\begin{document}

\title{{\bf Interference and the lossless lossy beam splitter}} 

\author{\large{JOHN JEFFERS}\\
\\
Department of Physics and Applied Physics,\\
University of Strathclyde,\\
107 Rottenrow,\\
Glasgow G4 0NG, UK.\\
email: {\tt john@phys.strath.ac.uk}\\
\\ }
\date{}
\vskip 3cm
\maketitle
\maketitle
\baselineskip=24pt
\vskip 1cm
{\bf Abstract. }
By directing the input light into a particular mode it is possible to
obtain as output all of the input light for a beam splitter that is
50\% absorbing. This effect is also responsible for nonlinear quantum
interference when two photons are incident on the beam splitter.
\vfill
\newpage

\section{Introduction}
The beam splitter is the main component of many optical
interferometers, both classical and quantum \cite{MW,HBT}.  Much of its
usefulness in quantum optics is derived from the fact that an
unentangled input beam can emerge from the device in an entangled
state. This state can then either have its quantum correlations
confirmed by measurement, or can itself be used in interference
experiments. Two photon interference is a higher order quantum
interference effect which has received much attention
\cite{HOM,R+T,SKC,L}.  In the classic experiments two identical photons
from different input arms falling simultaneously upon a 50/50 beam
splitter both exit the beam splitter from the same output arm; whereas
if the photons arrive at the beam splitter in the same input they are
distributed between the output arms as if they obeyed the laws of
classical probability, with each having a transmission and reflection
probability of $1/2$.

The relations between the input and output single-mode boson operators
for the different arms of a lossless beam splitter (Fig. 1) are
\cite{PSM,OHM,FL,CST}
\bea
\nn \hat{a}_{out} &=& t \hat{a}_{in} + r \hat{b}_{in},\\
\hat{b}_{out} &=& t \hat{b}_{in} + r \hat{a}_{in},
\eea where $t$ and $r$ are the transmission and reflection
coefficients. 
\begin{figure}[tb]
\infig{24em}{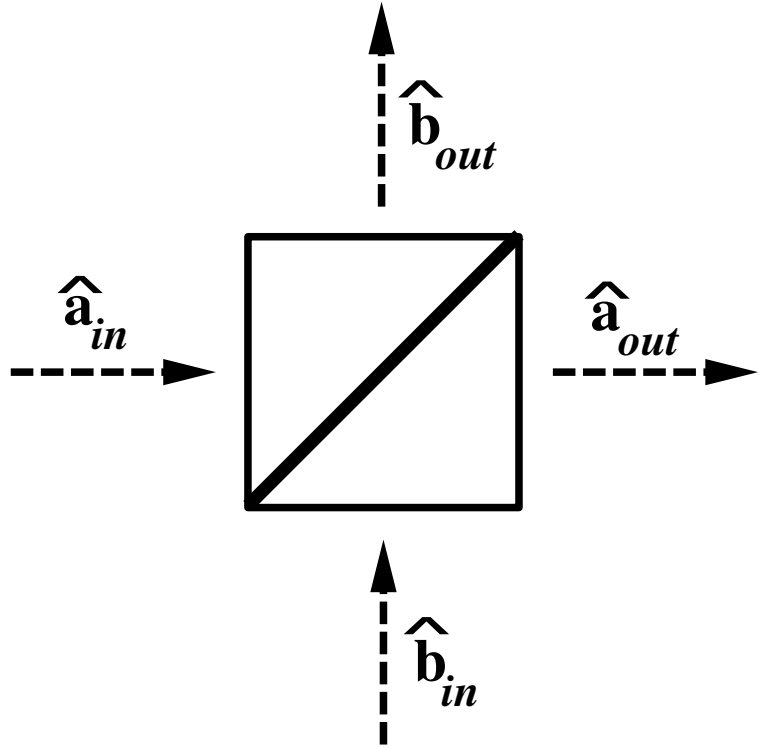}
\caption{The optical beam splitter. Two input fields are partially
reflected, partially transmitted into two output fields. In the lossy
case there are further unobservable noise inputs and outputs, which are
internal to the beam splitter.}
\label{fig:fig1}
\end{figure}
Throughout this paper discrete-mode theory is used. It is
merely an approximation to the more correct continuous-mode formalism,
but it is an adequate description here, provided that the reflection
and transmission coefficients can be treated as constants at the
frequencies of interest. Conservation of the boson commutators and the
requirement that independent inputs give independent outputs force the
coefficients to satisfy $|t|^2 + |r|^2=1$ and $t=\pm ir$. It is useful
to define superposition modes, the symmetric and antisymmetric
combinations of $\hat{a}$ and $\hat{b}$.  The annihilation operators
for these are (for either inputs or outputs)
\bea
\hat{c}=\frac{\hat{a} +\hat {b}}{\sqrt{2}}, \hat{d} =
\frac{\hat{b}-\hat{a}} {\sqrt{2}},
\eea so that the input and output operators are related by 
\bea
\hat{c}_{out} = (t+r) \hat{c}_{in}, \hat{d}_{out} = (t-r) \hat{d}_{in}.
\eea
It is clear from the above relations that these superposition operators
are not mixed by the beam splitter. The action of the device on any
input state which is factorisable into a product state of the
superposition modes, $|\psi_{in} \rangle = |\psi_c\rangle \otimes
|\psi_d \rangle$, is simply to produce a phase shift. For example, the
photon number state product
\bea
|\psi_{in} \rangle = |n_c\rangle \otimes |m_d \rangle \Rightarrow 
|\psi_{out} \rangle = \exp{[\pm i \theta(n-m)]}|n_c\rangle \otimes
|m_d \rangle,
\eea where $\cos \theta = t$ and $|\sin \theta | =r$. For the 50/50 beam
splitter two simple examples are the one photon superpositions
\bea
\nn |1_+\rangle &=& |1_c,0_d\rangle = \frac{1}{\sqrt{2}}
(|1_a,0_b\rangle + |0_a,1_b\rangle),\\
|1_-\rangle &=& |0_c,1_d\rangle = \frac{1}{\sqrt{2}} (|1_a,0_b\rangle -
|0_a,1_b\rangle).
\eea

It is also possible to construct states for which the beam splitter is
transparent with two photons.  Consider an input in one of the
superpositions 
\bea 
\label{2pm}
|2_\pm\rangle = \frac {1}{\sqrt{2}} \left(
|2_a,0_b\rangle \pm |0_a,2_b\rangle \right).
\eea Two photons falling on a beam splitter from the same input arm are
usually distributed between the outputs according to the laws of
classical probability but this is not the case here.  Photons in the
positive superposition are constrained to take different output arms.
This is an illustration of the reversibility of light beams in optics,
as it is merely the standard two photon interference phenomenon with
the inputs and outputs reversed. The input state in this case is not
factorisable in terms of states of the $c$ and $d$ modes. The negative
superposition, on the other hand, is a two photon example of a state
which passes through the beam splitter unchanged. This state can also
be written
\bea
\label{2-}
|2_-\rangle = \frac{1}{2} (\hat{a}^{\dagger 2} - \hat{b}^{\dagger 2})
|0_a,0_b> = -\hat{c}^\dagger \hat{d}^\dagger |0_c,0_d\rangle.
\eea The reason that the state is passed is now clear. It is
factorisable into a product state of the superposition modes.

Much of the theory of the beam splitter has been done assuming that it
is a lossless component \cite{PSM,OHM,FL,CST}, for which absorption is
simply an experimental problem that hinders the observation of
interesting effects. Absorption is unavoidable at some frequencies,
however, as optical devices must be compatible with the Kramers-Kronig
relations.  In recent years loss has come to be seen as a quantum
property in its own right, with its own effect on
correlations\cite{JB}. There has been some work on loss in fibre
couplers \cite{LBK} but it is only recently that a quantum theory of
the absorbing beam splitter was proposed \cite{BJGL}. All of the
features of the lossless device are retained but there are additional
quantum effects such as nonlinear absorption which are only possible
when loss is present.

In this paper a discrete-mode version of the lossy beam splitter theory
is used to show several interesting effects. Firstly, the beam splitter
may be rendered transparent or opaque, depending on the phase relation
of the transmission and reflection coefficients and the state of the
input light; and secondly, the nonlinear absorption effect discussed in
reference \cite{BJGL} is shown to occur with other two photon input
states.

\section{The lossy beam splitter}

In this section the theory of the lossy beam splitter \cite{BJGL} will
be outlined in discrete-mode form. The device relates two inputs to two
outputs as before (Fig. 1), but the presence of absorption forces
changes to the theory. The beam splitter input and output operators are
related by
\bea
\nn \hat{a}_{out} &=& t \hat{a}_{in} + r \hat{b}_{in} + \hat{f}_a,\\
\hat{b}_{out} &=& t \hat{b}_{in} + r \hat{a}_{in} +\hat{f}_b,
\eea where in this case $|t|^2 + |r|^2 < 1$. The noise operators,
$\hat{f}_a$ and $\hat{f}_b$, are necessary to preserve the commutators
of the observable outputs. We will assume that the noise modes are
unexcited, and can thus make no contribution to output photocounts.
Furthermore, the reflection and transmission coeffecients are not
constrained by a phase relation similar to the one for the lossless
beam splitter (that is, $t\neq ir$ in general). The requirement that
the beam splitter is lossy for coherent inputs does, however, give
another relation which the reflection and transmission coefficients
must satisfy,
\bea
\label{coherent}
|t\pm r|<1.
\eea 
The superposition operators for the lossy beam splitter are simply the
sum and difference operators as before. The output operators are
appended by noise operators,
\bea
\label{suploss}
\nn \hat{c}_{out} &=& (t+r)\hat{c}_{in} + \hat{f}_c,\\
 \hat{d}_{out} &=& (t-r)\hat{d}_{in} + \hat{f}_d,
\eea where $\hat{f}_c$ and $\hat{f}_d$ are respectively the sum of and
difference between the noise operators $\hat{f}_a$ and $\hat{f}_b$. The
theory outlined above is sufficient for this paper.  A fuller account
of the general theory in a continuous mode formalism is given in
reference \cite{BJGL}.

\section{The lossless lossy beam splitter} 
Consider a lossy beam splitter with transmission and reflection
coefficients which are equal or opposite, $t=\pm r$. Such a phase
relation would not be allowed for the lossless beam splitter, but the
extra freedom given by the noise operators associated with absorption
means that it is an allowed choice.  There is, however, one restriction
imposed by equation (\ref{coherent}) which limits the moduli of $t$ and
$r$ to be less than or equal to 1/2. The result of this is at best a
25/25 beam splitter with 50\% loss. The effects described also occur
with reduced visibility for beam splitters with greater absorption than
50\% but it is this ideal case which is assumed henceforth. It is clear
that for this choice of phase relation one of the superposition modes
is passed completely and the other is fully absorbed. The
$|1_\pm\rangle$ states are the one-photon excitations of these
superposition operators, and Table 1 summarises the effect of the beam
splitter on these inputs. 
\begin{center}
Table 1\\
\vskip 0.5cm
\begin{tabular}{|l|l|l|}\hline
input & beam splitter & output \\ \hline\hline
$|1_+\rangle$ & $t=r$ & $|1_+\rangle$ \\ \cline{2-3}
	      & $t=-r$ & $|0\rangle$ \\ \hline
$|1_-\rangle$ & $t=r$ & $|0\rangle$ \\ \cline{2-3}
	      & $t=-r$ & $|1_-\rangle$ \\ \hline
\end{tabular}
\end{center}
{\small Table 1. One photon superposition input states and output states for
the 25/25 lossy beam splitter.}
\vskip 0.5cm
Clearly the photons are passed or absorbed with either unit or zero
probability.  They are not subject to the randomness normally
associated with an absorbing medium that plagues other input states.
Although the table illustrates the results with one photon states, the
effect is not specifically quantum in nature. The beam splitter will be
rendered either transparent or opaque to any excitation of the
superposition modes, even a classical excitation such as the coherent
state.  The device exhibits the remarkable property that even though
half of the input light is normally lost, the full intensity can be
obtained at the outputs if the input is in a particular mode.
Furthermore, it is relatively easy to put input light into a beam
splitter in one of the superposition modes. A Mach-Zehnder
interferometer with the second beam splitter as the lossy one could be
set up to do this for narrow band input light (Fig. 2).
\begin{figure}[tb]
\infig{24em}{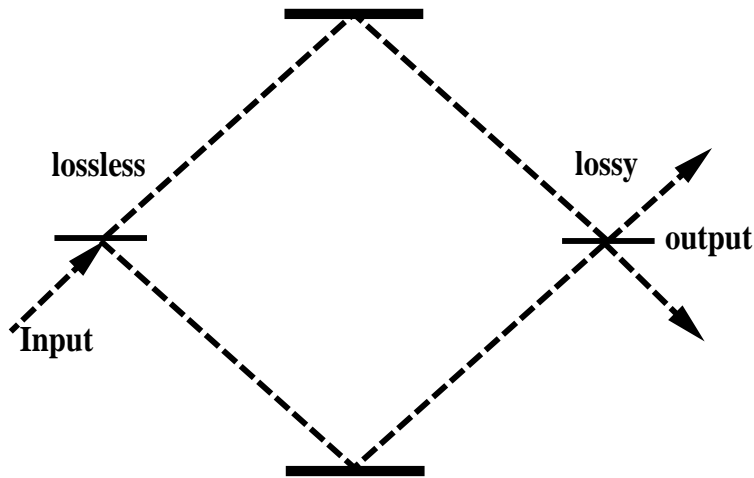}
\caption{Mach-Zehnder set up to obtain an input into the lossy beam
splitter in one of the superposition modes. The first beam splitter is a
standard lossless 50/50 device, the second is lossy, 25/25.}
\label{fig:fig2}
\end{figure}

\section{Two photon interference and nonlinear absorption}

The previous section described how a beam splitter can be made
transparent (or opaque) for classical or quantum excitations of the
superposition modes. In this section the effects of the absorbing beam
splitter on purely quantum input states are described. The standard two
photon interference experiment, in which a pair of photons incident on
a beam splitter from different input arms produce no coincidences at
the outputs, survives with reduced visibility in the lossy case
provided that the same phase relation $t=\pm ir$ is upheld. The same is
true of the extensions to two photon interference described in the
Introduction.  If, however, the beam splitter has equal or opposite $t$
and $r$, nonlinear absorption can take place. It has previously been
pointed out that for an input state with one photon in each input arm
it is not possible to obtain a one photon output \cite{BJGL}. Either
both photons pass through or both are absorbed. There are similar
nonlinear phenomena which occur when the input is in one of the
two-photon superposition states (\ref{2pm}). These are summarised in
Table 2. 
\newpage
\begin{center}
Table 2\\
\vskip 0.5cm
\begin{tabular}{|l|l|}\hline
input  & output ($t=\pm r$) \\ \hline\hline
$|1_a,1_b\rangle$ & mixture of $|0_a,0_b\rangle$ and
$\frac{1}{\sqrt{2}}(|1_a,1_b\rangle \pm |2_+\rangle)$\\ \hline
$|2_+\rangle$ & mixture of $|0_a,0_b\rangle$ and
$\frac{\pm 1}{\sqrt{2}}(|1_a,1_b\rangle \pm |2_+\rangle)$ \\ \hline
$|2_-\rangle$ &  $\mp|1_\pm\rangle$ \\ \hline
\end{tabular}
\end{center}
{\small Table 2. Two photon input states and output states for the 25/25 lossy
beam splitter.}
\vskip 0.5cm
Clearly nonlinear absorption takes place in each case. For the input
states specified in the first two lines of Table 2 the outputs cannot
consist of only one photon, whereas the input state in the third line
produces an output of one photon with certainty. In fact nonlinear
interference occurs for two photon input superpositions with any
relative phase angle (not merely $0$ or $\pi$ as in the Table 2).
There is a simple explanantion for the nonlinear interference in terms
of the superposition operators. The input state from the first line of
the Table 2 can be written
\bea
|1_a,1_b\rangle = \hat{a}^\dagger_{in} \hat{b}^\dagger_{in}
|0_a,0_b\rangle = \frac{1}{2} (\hat{c}^{\dagger 2}_{in} -
\hat{d}^{\dagger 2}_{in}) |0_c,0_d \rangle,
\eea when expressed in terms of the superposition operators for the
inputs.  One of $\hat{c}$ or $\hat{d}$ is completely absorbed by the
beam splitter, so that the part of the output which passes is either
$\hat{c}^{\dagger 2}_{out}/2$ or $-\hat{d}^{\dagger 2}_{out}/2$ acting
on the vacuum. The output state, found by re-expressing these operators
in terms of the standard output operators, is as given in Table 2.
Similarly the input state in the second line of Table 2,
$|2_+\rangle$, can be written $\frac{1}{2}(\hat{c}^{\dagger 2}_{in} +
\hat{d}^{\dagger 2}_{in}) |0_c,0_d \rangle$, so the explanation, and the
output state are the same (except for a simple phase change in one
case). The odd state out is the input in the third line of the
Table 2, which from equation (\ref{2-}) is $-|1_c,1_d\rangle$. There is
therefore one photon in each of the input superposition modes, one of
which is absorbed and the other passes. Hence the output must contain
one photon in the passed superposition mode.

\section{Conclusions}
In this paper it has been shown that the amount of absorption produced
by a lossy beam splitter can be reduced if the transmission and
reflection coefficients have an appropriate phase relation and the
input light is directed into particular modes. For a 50\% lossy beam
splitter, which ordinarily would absorb half of the light, it is possible
to obtain all of the input light in the output, thus rendering the device
transparent.

The modes which are affected are the superposition modes, the sum and
difference of the two inputs. One photon excitations of these modes are
quantum states which are useful in quantum information \cite{BEHPZ}.
The lossy beam splitter described here could therefore be used as a
filter, absorbing merely that part of the input which is not in the
required superposition.

A further consequence of the different amount of loss in each
superposition is the possibility of nonlinear absorption from a linear
device. For some two photon inputs the chance of a one photon output
can be eliminated. On the other hand for the two photon input which is
factorisable in terms of superposition input states, a single photon
output is obtained with certainty. It is clear from the above that, in
both the classical and quantum cases, interference is not simply washed
out by absorption.  Loss can enhance interference, and it even has
effects of its own which are not present in lossless environments.

\section*{Acknowledgments}
This research was supported by the Engineering and Physical Sciences
Research Council. I would like to thank Steve Barnett for useful
comments.

\end{document}